# Six-pack off-axis holography

MORAN RUBIN, GILI DARDIKMAN, SIMCHA K. MIRSKY, NIR A. TURKO, NATAN T. SHAKED*

*Department of Biomedical Engineering, Faculty of Engineering, Tel Aviv University, Tel Aviv 69978, Israel*
*Corresponding author: nshaked@tau.ac.il*

**We present a new holographic concept, named six-pack holography (6PH), in which we compress six off-axis holograms into a multiplexed off-axis hologram without loss of magnification or resolution. The multiplexed hologram contains straight off-axis fringes with six different orientations, and can be generated optically or digitally. We show that since the six different complex wave fronts do not overlap in the spatial-frequency domain, they can be fully reconstructed. 6PH allows more than 50% improvement in the spatial bandwidth consumption when compared to the best multiplexing method proposed so far. We expect the 6PH concept to be useful for a variety of applications, such as field of view multiplexing, wavelength multiplexing, temporal multiplexing, multiplexing for super-resolution imaging, and others.**

http://dx.doi.org/10.1364/optica.99.099999

Off-axis holography allows reconstruction from a single camera exposure, by inducing a small angle between the sample and reference beams creating the interference pattern of the hologram. This is possible since in the spatial frequency domain, there is a full separation between the auto-correlation terms and each of the cross-correlation terms, each of which contains the complex wave front of the sample. This separation is typically across a single axis, which allows compressing more information on the other axes as well. This enables optical multiplexing of several holograms within a single multiplexed hologram and full reconstruction of the data. Each hologram can contain additional data of the imaged sample, meaning that multiplexing allows recoding more information with the same amount of camera pixels. This can be beneficial for highly dynamic samples.

We have previously presented the technique of off-axis interferometry with doubled imaging area (IDIA) [1–3], in which we optically compress several off-axis interferometric images into a single camera image and gain extended field of view. In this case, the creation of the multiplexed hologram, recorded by the camera in a single exposure, is done by an external interferometric module, projecting onto the camera one reference beam and two or three sample beams at once. Other works used off-axis hologram multiplexing to compress two polarization states [4], two wavelengths [5,6], phase data and fluorescence [7], and to obtain super-resolution with synthetic aperture configuration [8].

Furthermore, even if the holograms are not projected onto the camera at once, digital off-axis hologram multiplexing has been shown to be beneficial for speeding up hologram reconstruction [9,10] (since only a single digital Fourier transform is needed to obtain the spatial frequency domain for isolation of the cross-correlation terms); and for compression [11] (since the multiplexed hologram can be sent to a distant point in a compressed manner and then be uncompressed there.

Regular off-axis holography uses a single sample beam and a single reference beam, which, in the spatial frequency domain, results in two auto-correlation terms located around the origin and two complex-conjugated cross-correlation terms located on both sides of the spatial frequency domain. Assuming that the maximum spatial frequency of the sample wave is $\omega_c$ on both axes, each of the cross-correlation terms occupies a spatial bandwidth capacity of $[-\omega_c, \omega_c]$, and the auto-correlation terms occupy a spatial bandwidth capacity of $[-2\omega_c, 2\omega_c]$ [1]. This is shown in Fig. 1(a). To avoid an overlap between the cross-correlation terms and the auto-correlation terms, the center of the spatial-frequency contents of the cross-correlation terms is shifted to at least $\pm 3\omega_c$ by adjusting the off-axis angle between the reference and sample beams, which requires a total spatial bandwidth of at least $8\omega_c$. However, for being efficient in using the spatial bandwidth of the camera and not wasting camera pixels, we need to use exactly $8\omega_c$. In other word, as also shown in Fig. 1(a), if the hologram and thus its spatial frequency plane have $N \times N$ pixels, the auto-correlation terms occupy $N/2 \times N/2$ pixels, and each of the cross-correlation terms occupies $N/4 \times N/4$ pixels [1]. In this case, the cross-correlation terms occupy 9.8% of the spatial frequency plane.

Tahara at el. [12] have proposed to position the cross-correlation terms on the diagonal axis, so that they can become larger, or even such that half of them will be positioned at the edge of the spectrum and appear on the other side. In this case, the cross-correlation terms occupy up to 24% of the spatial frequency plane.

The off-axis holographic encoding typically creates an empty space in the spatial-frequency domain, into which we can insert other cross-correlation terms encoding additional information on the sample. For example, Fig. 1(b) shows a multiplexed hologram with two orthogonal fringe directions that create two fully separable cross-correlation pairs in the spatial-frequency domain [4,5,6,9]. In this case, the cross-correlation terms occupy 19.6% of the spatial frequency plane. Till now, it was accepted that one can multiplex up to four off-axis image holograms without overlap in the spatial frequency domain, as shown in Fig. 1(c) [3,11,13]. In this case, the cross-correlation terms occupy 39% of the spatial frequency plane. In the current letter, we show, for the first time to our knowledge, two options of multiplexing six off-axis holograms without overlap in the spatial frequency domain. This principle is termed as six-pack holography (6PH).

Two options for 6PH are shown in Figs. 1(d) and (e). As shown in these figures, it is possible to position six cross-correlation terms in the spatial frequency domain without overlapping with each other and without overlapping with the auto-correlation terms. In both options, the total spatial frequency bandwidth capacity is $[-4\omega_c, 4\omega_c]$, and each cross-correlation term occupies spatial frequency bandwidth capacity of $[-\omega_c, \omega_c]$. In Fig. 1(e) only, due the cyclic property of discrete Fourier transform, each cross-correlation term that moves out of the spatial Fourier domain appears in the other side of this domain, on both axes.

Figure 2 presents a scheme of the entire multiplexing process. In Step 1, we acquire off-axis holograms on the digital camera. In Step 2, we multiplex the six holograms. If optical multiplexing is performed, six sample and reference beam angle combinations are projected onto the digital camera at once, so that each combination creates straight off-axis fringes in a different direction, and each fringe direction yields a different cross-correlation pair in the spatial frequency domain that does not overlap with any of the other five cross-correlation pairs. Alternatively, one way to perform digital multiplexing is to acquire six regular off-axis holograms, spatially filter their six cross-correlation terms, and digitally position them in the spatial frequency domain to create the multiplexed off-axis hologram by an inverse Fourier transform.

In both scenarios, the multiplexed hologram encodes six complex wave fronts, but it contains real numbers only (representing intensity of an interference pattern) and no complex numbers, and it can be sent or stored in its compressed version when needed. 6PH allows the cross-correlation terms to occupy 59% of the spatial frequency plane; thus more than 50% improvement compared to four hologram multiplexing. We believe that 6PH presents the optimized camera space-bandwidth product usage for off-axis holography that allows optical off-axis hologram multiplexing where the auto-correlation terms are simultaneously acquired.

Step 3 of Fig. 2 presents the reconstruction of the six complex wave fronts encoded into the multiplexed hologram. This reconstruction step can be performed digitally in the computer, or optically by displaying the multiplexed hologram on a spatial light modulator and projecting the six reconstructed wave fronts.

To demonstrate the proposed technique, we have acquired red blood cells during flow and digitally multiplexed each group of six holograms from the sequence into a single dynamic multiplexed hologram containing the entire data but in sixth of the temporal resolution of the original data. This specific application can be useful for visualization purposes, for example, if the displaying device allowing optical reconstruction or the observer viewing the optical reconstruction has lower temporal resolution capabilities than the camera acquiring the holograms.

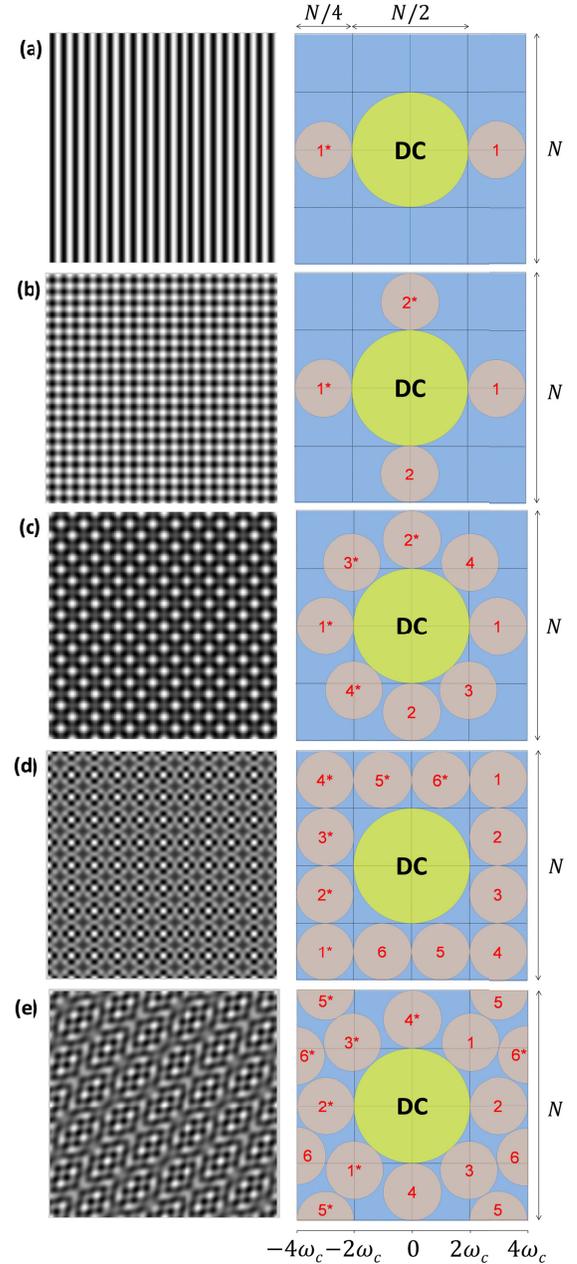

Fig. 1. Schematic illustrations of off-axis holograms (left) and the coinciding spatial frequency domains (right) for: (a) Standard off-axis holography. (b) Multiplexing two off-axis holograms with orthogonal fringe directions. (c) Multiplexing four off-axis holograms (the current paradigm for optimized space-bandwidth product). (d) First option for multiplexing six off-axis holograms. (e) Second option for multiplexing six off-axis holograms. In the spectra images, DC denotes the auto-correlation terms. The numbered circles around it denote the cross-correlation term, where the coinciding complex conjugate cross-correlation terms are denoted by a number and an asterisk.

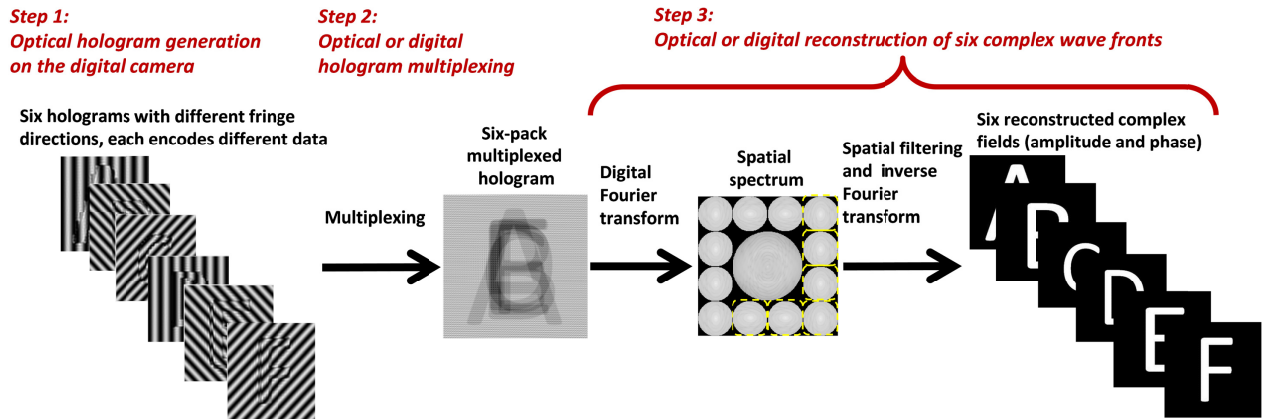

Fig. 2. Schematic illustration of the 6PH multiplexing method (Option 1 from Fig. 1(d) is shown).

Our experimental setup for capturing the cells during flow includes a 10 mW, 632 nm Helium-Neon laser that illuminates a transmission microscope. The laser beam passes through the sample and is magnified by 50×, 0.55 NA microscope objective (M Plan Apo, Mitutoyo). Before the image plane of the system, where a digital camera is positioned [DCC1545M, Thorlabs], we positioned our external flipping interferometric module, which is based on a modified Mach-Zehnder interferometer [14]. This interferometer flips half of the optical field of view on top of the other half using a retro-reflector. If half of the optical field is empty, we effectively use this half as the reference beam and the other half as the sample beam. The two beams interfere on the digital camera at an angle, creating an off-axis image hologram. To meet the requirement of half empty optical field of view, we positioned the optical field of view on the side of the microfluidic channel (μ-slide VI0.1, IBIDI), so that only half of the optical field of view contains flowing cells and the other one contains the empty channel glass [14].

As shown in Fig. 3(a) and Visualization 1, 720 off-axis holograms were recorded during the red blood cells flow in the channel. Figure 3(b) and Visualization 2 shows the quantitative phase map obtained by reconstructing this hologram using the typical off-axis holography phase extraction algorithm, containing spatial filtering of the cross-correlation term and phase unwrapping to solve $2\pi$ ambiguities [9]. Then, one of the cross-correlation terms of each hologram was extracted using digital Fourier transform, and each six cross-correlation terms from the hologram sequence were put into the Fourier domain of a single multiplexed hologram, as shown in Fig. 4(a). By inverse Fourier transform, we then obtain the dynamic six-pack multiplexed hologram shown in Fig. 4(b) and Visualization 3. Although this multiplexed hologram encodes six complex wave fronts from six temporal events of the dynamic sample, it contains real values only. This multiplexed hologram can therefore be stored or sent in its compressed form, and/or be reconstructed optically with decreased temporal resolution. This can be done optically (e.g., by projecting the multiplexed hologram on a spatial light modulator) or digitally. We reconstructed each of the wave fronts compressed into multiplexed hologram digitally by Fourier transforming it and cropping the six relevant cross-correlation terms. Then, each of these cross-correlation terms was inversed Fourier transformed and the quantitative phase of the result was retrieved using $2\pi$ phase unwrapping [9]. Figure 4(c) and Visualization 4 show the six reconstructed quantitative phase profiles, each of which is at sixth of the frame rate of original quantitative phase profile shown in Fig. 3(b), and encodes different temporal events.

In SPH, despite the fact that there is no overlap in the spatial frequency domain, yielding the possibility of reconstruction of six different cross-correlation terms from a single multiplexed hologram, one should note that there is a risk for loss of information due to the fact the six holograms share the same dynamic range of the camera grayscale levels. This is valid only in case we perform the multiplexing optically by projecting onto the digital camera six sample and reference beam combinations, using one of the optical multiplexing methods for off-axis holograms presented in Refs. [1-6]. To examine this possible quality loss, before reconstructing the multiplexed hologram, we decreased the dynamic range of the hologram into 8 bits (256 grayscale levels), which was the dynamic range of our camera. After this bit depth decrease, no visual changes in the reconstruction quality were seen in comparison to the reconstruction shown in Fig. 3(b) and Visualization 2. The mean square error (MSE) between the phase reconstructed from the multiplexed hologram after 8-bit dynamic range limit and the phase reconstruction from the regular hologram was 0.3%, showing minimal quality decrease. This is valid for phase object, as demonstrated here. However, the problem of quality loss due to grayscale dynamic range sharing is expected to be more severe when the amplitude of the sample becomes non-negligible [15]. Solving this problem might require cameras with larger bit depth, dependent on the application.

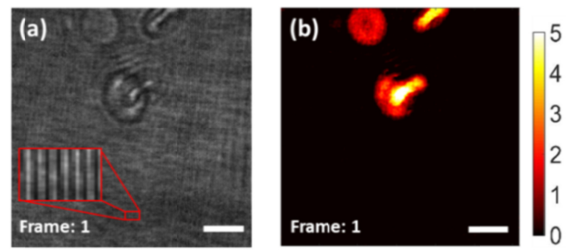

Fig. 3. Quantitative phase imaging of red blood cells in flow. (a) Dynamic off-axis hologram (Visualization 1, 48 frames per second). (b) Reconstructed dynamic phase profile (Visualization 2). Colorbar represents phase values in radians.

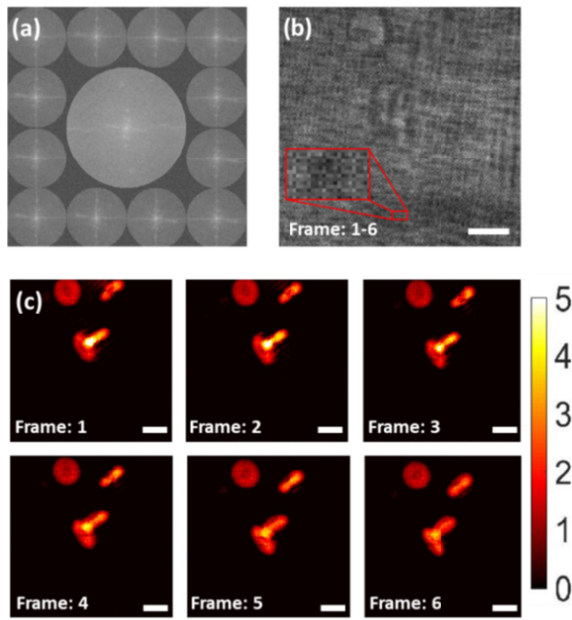

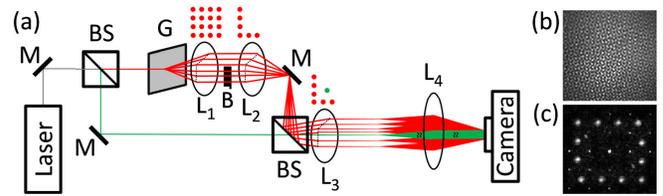

Fig. 4. (a) The power spectrum of the multiplexed hologram of the first six frames. (b) Dynamic multiplexed hologram at sixth of the frame rate of the original dynamic hologram (Visualization 3). (c) Six dynamic reconstruction phase profiles (Visualization 4), each at sixth the frame rate of the original dynamic phase profile. White scalebars represent 5 µm on the sample. Colorbar represents phase values in radians.

In this Letter, we demonstrated digital hologram multiplexing of experimental holographic data; thus, we first acquired regular off-axis hologram and then compressed six holograms into a single one, which is useful for displaying fast dynamics. In this case, we can use the full spatial frequency spectrum without creating unwanted cross-terms which may occupy useful space in the spatial frequency spectrum. A possible setup for optical hologram multiplexing is, for example, the external IDIA module previously proposed by us [1], but with six rotated retro-reflectors, rather than two, each creating a difference fringe orientation in the multiplexed hologram. Another option for such an optical multiplexing setup is shown in Fig. 5(a). This setup uses a specialized grating (DOC Optics Corp) that splits one of the beams in a Mach-Zehnder interferometer into 4×4 beam array, from which six beams arranged in an 'L' shape are selected. The other interferometric beam stays normal to the camera plane. This creates on the camera the optically multiplexed interference pattern shown Fig. 5(b), with the cross-correlation terms located at non-overlapping location in the spatial frequency spectrum (see Fig. 5(c)). When applying optical hologram multiplexing, one needs to avoid unwanted cross-terms between nonmatching pairs of beams. This can be done by using different wavelengths for each sample and reference beam pair, or using the low coherence of the source (such that the path difference between different pairs of beams is longer than the coherence length of the source).

To conclude, we have presented 6PH, a method for multiplexing six off-axis holograms into a single multiplexed hologram, where the multiplexing can be done optically or digitally. The multiplexing allows optimized usage of the spatial frequency domain by compressing six cross-correlation terms without overlap, and thus reconstruction of all six of them.

Fig. 5. (a) 6PH optical multiplexing system. BS, beam splitter. G, 1:4×4 grating. M, mirror. B, beam block. $L_1$, $L_2$, 1:1 telecentric lens system. $L_3$, $L_4$, 1:5 telecentric lens system. (b) The resulting 6PH multiplexed (sample-free) interference pattern. (c) The coinciding power spectrum.

6PH allows a great improvement of the space-bandwidth product of more than 50% compared to the best method previously proposed, and we believe it represents the optimal spatial bandwidth consumption for optical hologram multiplexing. We have demonstrated 6PH for decreasing temporal resolution of dynamic hologram recording by obtaining a digitally compressed multiplexed hologram containing six complex wave fronts from six temporal events of the samples, but at sixth of the original temporal resolution and still without losing temporal events. The decrease in the temporal resolution in the multiplexed hologram is straightforward and does not require adjusting the temporal interval captured in each sub-hologram to the framerate of the camera. This digital multiplexing implementation is useful for real-time display of fast events, which the observer or the hologram displaying device cannot handle with. In general, the 6PH principle is expected to be useful for a wide variety of applications, with focus on compressing of holographic data and quantitative imaging of fast dynamics, including extended field of view imaging, multi-color imaging, multi-polarization imaging, super-resolution imaging, tomographic phase microscopy data, and others.

**Funding.** Horizon 2020 European Research Council (678316).